\documentclass[10pt]{article}
 \usepackage{pvm2col}
\usepackage{amsmath,amssymb,array,graphicx,psfrag}

\usepackage{empheq}

\usepackage{mathcomp}  
\usepackage{textcomp}  

\input cyracc.def
\font\tencyr=wncyr10

\everymath{\displaystyle}

\numberwithin{equation}{section}

\newcommand{\eq}{Eq.~}   
\newcommand{\Eq}{Equation~}
\newcommand{\eqs}{Eqs.~}

\newcommand{\fig}{Fig.\:}
\newcommand{\Fig}{Figure~}

\newcommand{\sliver}{{\hskip0.6pt}}  
\newcommand{\ssliver}{{\hskip0.9pt}}  
\newcommand{\Sliver}{{\hskip1.2pt}}  
\newcommand{\antisliver}{{\hskip-0.6pt}}

\newcommand{\Antisliver}{{\hskip-1.2pt}}

\newcommand{\am}{angular momentum}
\newcommand{\ahm}{angular-momentum}
\newcommand{\aam}{absolute angular momentum}

\newcommand{\azm}{absolute zonal momentum}
\newcommand{\pv}{po\-ten\-tial vor\-ti\-ci\-ty}

\newcommand{\qg}{quasigeostrophic}
\newcommand{\sw}{shallow water}
\newcommand{\shw}{shallow-water}
\newcommand{\gm}{generalized mixing}
\newcommand{\ghm}{generalized-mixing}  
\newcommand{\rf}{redistribution function}
\newcommand{\bdf}{bulk displacement function}  
\newcommand{\jss}{jet self-sharpening}

\newcommand{\jhrs}{jet-resharpening}

\newcommand{\q}{q}
\newcommand{\qi}{\q_{\ssliver i}}
\newcommand{\ql}{\q_{\rm{\ell}}}
\newcommand{\qbar}{\bar{\q}}
\newcommand{\qlbar}{\qbar_{\rm{\ell}}}

\newcommand{\topogfull}{b}  
\newcommand{\topogmean}{\bar\topogfull}
\newcommand{\topogfluct}{\tilde\topogfull}
\newcommand{\y}{y}
\newcommand{\x}{x}
\newcommand{\timevar}{t}      
\newcommand{\xx}{x'}

\newcommand{\aaa}{\sliver a \sliver} 
\newcommand{\Q}{Q}
\newcommand{\Qi}{Q_{\ssliver i}}

\newcommand{\dd}{d}

\newcommand{\yy}{y'}
\newcommand{\yyy}{y''\antisliver}
\newcommand{\lims}{_{\!-\!L}^{L}}
\newcommand{\intd}{\int\lims \!\!}
\newcommand{\densintd}{\dens H \!\! \int\lims \!\!}
\newcommand{\Domain}{\mathcal{D}}
\newcommand{\intxy}{\int \!\!\!\! \int_{\Domain} \!\!}
\newcommand{\dy}{\, d\y}
\newcommand{\dyy}{\, d\yy}
\newcommand{\dyyy}{\, d\yyy}
\newcommand{\dx}{\, d\x}
\newcommand{\dxx}{\, d\xx}

\newcommand{\xyxy}{(\xx\!,\yy;\x,\y)}
\newcommand{\R}{R}

\newcommand{\Ryy}{\R(\yy,\y)}
\newcommand{\Ryyy}{\R(\yy\antisliver,\y\sliver;\sliver\y_0)}

\newcommand{\Rf}{\R_{2}}   
\newcommand{\Rs}{\R_{1}}   
\newcommand{\RsRf}{\Rf \Antisliver \circ \Antisliver \Rs\Sliver}
\newcommand{\Rd}{\Delta\antisliver\R}      
\newcommand{\Rdyy}{\Rd(\yy,\y)}
\newcommand{\Rdyyy}{\Rd(\yy\antisliver,\y\sliver;\sliver\y_0)}  

\newcommand{\D}{[-L,L]}
\newcommand{\Mabs}{M}
\newcommand{\Mvar}{\tilde{\Mabs}}  

\newcommand{\Mvari}{\Mvar_{i}}
\newcommand{\Mvarl}{\Mvar_{\rm{\ell}}}

\newcommand{\rfwd}{s}

\newcommand{\Etabig}{\mbox{\tencyr I}}  
\newcommand{\Etamod}{\;\widehat{\phantom{|}}\hspace{-5.5pt}\mbox{\tencyr I}}
\newcommand{\etamod}{\hat\eta}

\newcommand{\G}{G}   

\newcommand{\dens}{\rho_{0}}

\newcommand{\qf}{\q_{1}}
\newcommand{\qfxy}{\qf(\xx,\yy)}
\newcommand{\qs}{\q_{2}}
\newcommand{\qsxy}{\qs(\xx,\yy)}

\newcommand{\qm}{\q_{2}}
\newcommand{\ym}{\y_{2}}

\newcommand{\uoverline}{\bar u}

\newcommand{\bigjump}{\q_{s}}
\newcommand{\littlejump}{\delta  \bigjump}
 
\newcommand{\ydumm}{\tilde y}  

\newcommand{\delM}{\Delta\antisliver\Mabs}  
\newcommand{\delq}{\Delta\q}  
\newcommand{\deluoverline}{\Delta\antisliver\uoverline}  
\newcommand{\LD}{L_{D}}

\newcommand{\newpartialmetric}{{\sliver Y\Antisliver}}  

\newcommand{\sgn}{{ \rm sgn }}

\newcommand{\functional}{\scriptstyle (\cdot) \displaystyle}

\newcommand{\comment}[1]{}

\newcommand{\andrefs}{and references therein}

\renewcommand{\forall}{\mbox{for all}}

\newcommand{\myabstract}{
An initial
zonally symmetric
 \qg\ \pv\ (PV) distribution
$\qi(\y)$ is 
subjected to complete or partial mixing within
some finite zone $|\y|<L$,  
where
$\y$
is latitude.  
The change in $\Mabs$, the total
\aam, between  
the initial and any later
time is considered.
For standard \qg\ \shw\  
beta-channel dynamics  
it is proved that, 
for any $\qi(\y)$ such that
$\dd\qi/\dd\y > 0$ 
throughout  
$|\y|<L$,  
the change in
$\Mabs$ is always negative.
This theorem holds even when
``mixing''  
is understood in  
the most general possible  
sense.  
Arbitrary stirring or  
advective rearrangement is included,
combined   
to an arbitrary extent with  
spatially inhomogeneous diffusion.  
The theorem holds  
whether or not  
the PV distribution
is zonally symmetric 
at the later time. 
The same theorem governs  
Boussinesq potential-energy changes due to buoyancy mixing in
the vertical.
For 
the standard \qg\ beta-channel dynamics 
to be valid
the Rossby deformation length
$\LD \gg \epsilon L$
where $\epsilon$ is the
Rossby number;  
when $\LD=\infty$ 
the theorem applies not only to
the beta-channel, 
but also to a single barotropic
layer on   
the full  
sphere,
as considered in the recent work of 
\citeauthor*{Dunkerton:2008}
on ``PV staircases''. 
It follows that
the $\Mabs$-conserving PV 
reconfigurations studied
by   
those authors 
must 
involve 
processes describable as PV 
unmixing,  
or antidiffusion, 
in the sense of time-reversed diffusion.
Ordinary \jss\ and jet-core acceleration do not, by  
contrast, require unmixing,  
as is shown 
here  
by detailed analysis.   
Mixing in the jet flanks suffices.
The theorem
extends  
to multiple layers and
continuous stratification. 
A least upper bound 
and 
greatest lower bound
for the change in $\Mabs$
is obtained for
cases in which $\qi$ is neither
monotonic nor zonally symmetric.
A corollary is a 
new nonlinear stability theorem for shear flows. 
}
\begin{document}
%
%%%%%%%%%%%%%%%%%%%%%%%%%%%%%%%%%%%%%%%%%%%%%%%%%%%%%%%%%%%%%%%%%%%%%
% TITLE
%%%%%%%%%%%%%%%%%%%%%%%%%%%%%%%%%%%%%%%%%%%%%%%%%%%%%%%%%%%%%%%%%%%%%
\title{\textbf{\large{A general theorem on \ahm\ changes
       due to potential vorticity mixing
       and on potential-energy changes
       due to buoyancy mixing 
}}}
%
% Author names, with corresponding author information. 
% [Update and move the \thanks{...} block as appropriate.]
%
\author{\textsc{Richard B. Wood}  
                                \thanks{\textit{Corresponding author address:} 
                                Richard B. Wood,
                                University of Cambridge, Department of Applied Mathematics and 
                                Theoretical Physics, Wilberforce Road, Cambridge CB3 0WA, UK
                                \newline{E-mail:
latex@ametsoc.org}}\quad\textsc{and Michael E. McIntyre}\\   
\textit{\footnotesize{University of Cambridge, Department of Applied Mathematics and 
                      Theoretical Physics, Cambridge CB3 0WA, UK}}
}

% The following block of code will handle the formatting of the title page depnding on whether
% we are formatting a double column (dc) author draft or a single column paper for submission.
% AUTHORS SHOULD SKIP OVER THIS... There is nothing to do in this section of code.
\ifthenelse{\boolean{dc}}
{
\twocolumn[
\begin{@twocolumnfalse}
\amstitle
\vspace{-1em}

% Start Abstract (Enter your Abstract above.  Do not enter any text here)
\begin{center}
\begin{minipage}{13.0cm}
        \begin{center}
					In press with \emph{J. Atmos. Sci}, 
					vol.\ \textbf{67} (2010). 
					\\
					Submitted 19 August 2009, 
					accepted 9 November 2009. 
        \end{center}
        \vspace{0.3em}
\begin{abstract}
        \myabstract
        \newline
        \begin{center}
                \rule{38mm}{0.2mm}
        \end{center}
\end{abstract}
\end{minipage}
\end{center}
\end{@twocolumnfalse}
]
}
{
\amstitle
\begin{abstract}
\myabstract
\end{abstract}
\newpage
}

%%%%%%%%%%%%%%%%%%%%%%%%%%%%%%%%%%%%%%%%%%%%%%%%%%%%%%%%%%%%%%%%%%%%%
% MAIN BODY OF PAPER
%%%%%%%%%%%%%%%%%%%%%%%%%%%%%%%%%%%%%%%%%%%%%%%%%%%%%%%%%%%%%%%%%%%%%

\section{Introduction}
\label{sec:intro}

\begingroup
\def\thefootnote{\fnsymbol{footnote}}
\footnotetext[1]{\textit{Corresponding author address:} 
                                Richard B. Wood,
                                University of Cambridge, Department of Applied Mathematics and 
                                Theoretical Physics, Wilberforce Road, Cambridge CB3 0WA, UK
                                \newline{\textit{E-mail:} R.B.Wood@damtp.cam.ac.uk}}
\endgroup

Ideas about the turbulent  
mixing of vorticity and \pv\ (PV),
going back to the pioneering work of
\citet{Taylor:1915,Taylor:1932}, 
\citet{Dickinson:1969},         
\citet{Green:1970}, and
\citet{Welander:1973}
are an important key to 
understanding  
such phenomena as
Rossby-wave 
``surf zones'',  
\jss,  
and eddy-transport barriers.  
For a review see \citeauthor{Dritschel:2008} (\citeyear{Dritschel:2008}, hereafter DM08); also,
for example,
\citet{Killworth:1985},
\citet{Hughes:1996},
\citet{Held:2001b},
\citet{McIntyre:2008},  
  \citet{Esler:2008b,Esler:2008},  
and
\citet{Buhler:2009}.
A key point
is
that PV mixing 
generically 
requires   
\ahm\ changes.
In the real world  
those
changes  
are usually mediated by,  
or catalyzed by, 
the radiation stresses or
Eliassen--Palm fluxes
due to Rossby waves and other wave types,
including the form stresses  
exerted across undulating stratification surfaces.
Usually, therefore, 
there is no such thing as
turbulence without waves. 

PV mixing
by baroclinic and barotropic shear instabilities
depends on radiation stresses internal
to the system, mediating
\ahm\ changes that add to zero.  
Cases like that of
Jupiter's  
stratified weather layer  
probably
depend on   
form  
stresses exerted from below, as
is known to be   
true of  
the terrestrial stratosphere. 

\begin{figure}
  \noindent\includegraphics[width=19pc,angle=0]{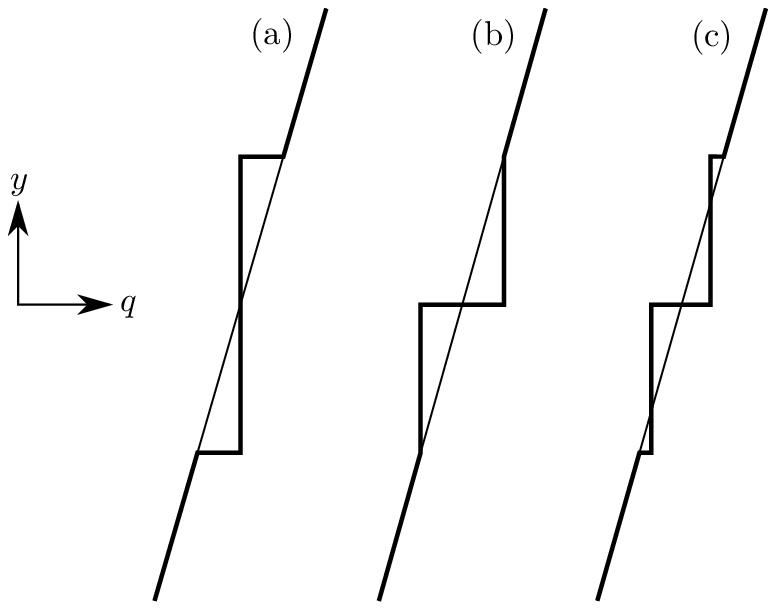}\\
\caption{\small  
Examples of 
initial and final
zonally symmetric
PV profiles (thin and thick lines respectively). 
For each initial profile the PV increases linearly with
latitude 
$\y$.
The examples could represent PV distributions in a \qg\ \sw\ system,
in a nondivergent barotropic system,  
or in a single layer within a
multilayered or continuously stratified system.
The \ahm\ changes $\delM$
are respectively negative, positive, and zero
in cases (a)--(c), 
all of which satisfy (\ref{eq:totalpv}). 
Cases (b) and (c)
require unmixing, or anti-diffusion.
\citet{Dunkerton:2008}
restrict attention to cases like (c).
}
\label{fig:mix-examples}
\end{figure}

Consider for instance the \qg\ thought experiment shown in
\fig\ref{fig:mix-examples}a.  
This is
an idealization of Rossby-wave surf-zone formation.  
An initially linear PV profile (thin line) is mixed
such that the PV
becomes uniform within a finite
latitudinal 
zone $|\y|<L$ (thick zigzag line).
The mixing is
assumed to be  
conservative in the sense that
\begin{equation}
\label{eq:totalpv}
\int\!\!\!\!\int \!\! \dx \! \dy \, \delq  = 0
\end{equation}
where $\q$ is the \qg\  
PV and $\delq$ its change
due to mixing;
$\dx \! \dy$ is the horizontal area element. 
It is well known that, according to standard \qg\ theory, the  
resulting change
$\delM$  
in the total 
\aam\ $\Mabs$ is negative or  
retrograde, in this  
special  
case 
with
the initial profile 
linear
in $\y$.\footnote{For
an explicit demonstration, see, e.g., DM08 \eqs(7.1)--(7.2)
and below \eq(A.4),  
noting that the
integration by parts 
at the penultimate step is valid
both for bounded and unbounded 
beta channels  
provided that the change $\deluoverline$ in the zonal-mean zonal flow
vanishes at the side boundaries 
\citep{Phillips:1954}.
For the unbounded channel, $\delM$ is
entirely due to the  
ageostrophic  
mass shift associated
with the northward residual
circulation,  
since
$\deluoverline$
integrates to zero.
For the bounded channel
there are contributions both from the mass shift and from
$\deluoverline$.
  }  
Such 
\ahm\ deficits
are key to understanding why, 
for instance,
breaking stratospheric Rossby waves
gyroscopically pump 
a Brewer--Dobson circulation that is
always poleward and never  
equatorward.  
The troposphere exerts a persistently
westward form stress on the stratosphere.
The physical reality of such
surf-zone formation events  
and their tendency to mix PV  
has been verified in a vast number of
observational and modeling   
studies, including   
studies of the stratospheric ozone layer
\citep[e.g.,][\andrefs]{Lahoz:2006}. 

In an interesting recent paper in this journal,  
\break
\citeauthor{Dunkerton:2008} (\citeyear{Dunkerton:2008}, hereafter DS08), 
consider a class of PV reconfigurations in a single layer
on the sphere,
with zonally symmetric
initial and final states, 
satisfying~(\ref{eq:totalpv}) and
constructed so as  
to make $\delM=0$. \
In DS08  
the dynamics is nondivergent barotropic.  
That is,
the 
Rossby deformation length $\LD=\infty$,
and $\q$ is
the absolute vorticity.
As 
illustrated  
in DS08,  
the constraint~(\ref{eq:totalpv})            
does not by itself dictate the sign of $\delM$.
However, in view of the ubiquity of radiation stresses
in real atmospheres and oceans, 
one is led to question
whether the assumption $\delM=0$ is
a natural one  
for realistic models.  

\Fig\ref{fig:mix-examples}b shows
a simple case where $\delM$ is positive
and \fig\ref{fig:mix-examples}c a case where $\delM$ is zero
as in DM08.  
Both these cases
must involve 
unmixing, or anti-diffusion.  
To go from the initial to the
final state in \fig\ref{fig:mix-examples}b
or \fig\ref{fig:mix-examples}c, 
one must transport $\q$ nonadvectively against its local gradient, 
at least in some locations (x,y). 
Such locally countergradient transport 
seems unnatural,  
at least
as a persistent phenomenon  
in a model                  
free of gravity wave stresses.  

To exclude 
such
countergradient transport  
we will
restrict the 
PV reconfigurations,
throughout this paper,
not only to respect~(\ref{eq:totalpv}) but also
to be describable as
``generalized partial mixing'', or  
``\gm'' for brevity.  
This
will be made precise in section~\ref{sec:gm},
using the standard
``mixing kernel'' or ``\rf''
formalism,  
but in essence means
that no unmixing is allowed.
With that restriction,
and a nonvanishing change
$\delq$  
in the PV profile,
we will prove
a theorem stating
that $\delM$ will 
always be negative, 
as it is in the special case  
of \fig\ref{fig:mix-examples}a, 
provided only that the initial
PV profile is 
zonally symmetric and monotonically
increasing  
in $\y$.
In all other respects the initial profile  
is arbitrary.

This theorem --- which we designate as ``basic"  
since it underpins the rest of our analysis ---
has been proved in several different ways.
In section~\ref{sec:basictheorem}  
we give what we think is the
most readable
of these proofs, 
after relating $\delM$ to $\delq$ in  
sections~\ref{sec:Mdef}
and~\ref{sec:other-systems}.
Section~\ref{sec:ape}
points out that 
the basic theorem has 
an alternative interpretation
in terms of potential energy and  
available potential energy.

Central to the proof 
in section~\ref{sec:basictheorem}
is an intrinsically nonnegative ``\bdf''
constructed from the \rf.  
Its physical meaning is  
briefly discussed in
section~\ref{sec:Eta-meaning}. \
Appendix~A 
presents one of the alternative proofs,  
based on a second, quite different nonnegative function.
That function is related to  
the so-called momentum--Casimir invariants
of Hamiltonian theory and therefore mathematically
related, also, to energy--Casimir invariants
\citep[e.g.,][]{Shepherd:1993}
again connecting with the
theory of available potential energy.
This second nonnegative function 
is constructed  
from the initial PV profile  
rather than from the \rf.

The upshot is that from
section~\ref{sec:basictheorem}
and
Appendix~A 
we have two entirely different proofs  
not only of the sign-definiteness of $\delM$,
but also of the sign-definiteness of the
potential energy change due to generalized
vertical mixing
of an initially  
stable stratification.
This  
generalizes classical results both on
vortex dynamics
  \citep{Arnold:1965} 
and on
available potential energy
\citep{Holliday:1981},  
beyond the Hamiltonian framework.  
The potential-energy interpretation applies to
a Boussinesq model with a
linear equation of state.  
The proofs in
sections~\ref{sec:basictheorem}   
and
Appendix~A 
provide us with two completely different
types of sign-definite integral
formulae, typified by  
(\ref{eq:DMhat3})
and
(\ref{eq:DMhat3-casimir})  
below,
for $\delM$ and for the analogous
sign-definite  
change in
potential energy.  

As summarized in section~\ref{sec:furtherimplications},
the 
basic theorem 
covers
three classes of model system: 
first, a \shw\ beta channel,
second, a
stratified \qg\ beta channel,
and third,
the system considered in 
DS08 --- a sphere with $\LD=\infty$.
Section~\ref{sec:furtherimplications}
also points out that the basic theorem 
provides, as a corollary,
a 
substantial 
generalization of   
the Charney--Stern   
shear-flow stability theorem,
related also to the classical work of \citet{Arnold:1965}.

Section~\ref{sec:genproof}
presents 
a generalization 
of the basic theorem 
to cases in which
the initial PV profile is neither
monotonic nor zonally symmetric.

Sections~\ref{sec:resharp}  
and~\ref{sec:genjet} 
discuss  
how the basic theorem applies to
\jss\ 
by PV mixing in the jet flanks.
In section~\ref{sec:resharp}  
we show via a specific example  
how a process
for which $\delM$ must always  
be negative can nevertheless result in jet-core acceleration.
Section~\ref{sec:genjet} goes  
on to prove a 
much more general result. 
For the \shw\ model,
PV  
mixing anywhere  
on one or both 
the flanks of a jet
must
always  
accelerate the jet core,
provided that the jet is zonally symmetric
both before and after mixing.

Section~\ref{sec:beyond} briefly  
discusses   
the possibility
of extending these results beyond 
\qg\  
to more accurate models.
So far, we have failed to find
such extensions.   
Obstacles to progress include the
nonlinearity  
of accurate PV inversion operators.
In the concluding remarks,
section~\ref{sec:conclu}, 
we touch on  
the implications for
models of geophysical turbulence.
In particular,
our results underline      
the need 
to pay closer  
attention to
the \ahm\ budget in such models.

\section{Definition of generalized mixing}
\label{sec:gm}

As well as ordinary  
diffusion-assisted mixing  
we want to include 
the limiting case of  
purely advective rearrangement,  
or pure stirring.
All such cases, from pure stirring to partial mixing
to perfect mixing,  
can be described as linear operations 
on the PV field.  
They are conveniently represented
in terms 
of a Green's  
function or integral
kernel   
in the standard way 
\citep[e.g.,][]{
Pasquill:1983,
Fiedler:1984,
Stull:1984,
Plumb:1988,
Shnirelman:1993,
Thuburn:1997,
Esler:2008b}.
Such Green's functions have properties akin to
probability density functions,
and are  
called
bistochastic or  
doubly stochastic.  
The corresponding linear operators are
sometimes  
called
polymorphisms.

The Green's function formalism is essentially the same
for all the
model systems 
under consideration,
including those describing 
potential-energy changes.  
So   
it will suffice to restrict attention at first to
the \shw\  
case. 
For a general
two-dimensional 
domain $\Domain$,
let $\qi(\x,\y)$ be the initial PV distribution
and  $\ql(\x,\y)$   
the PV distribution at some later time.
Because of linearity
and horizontal non-divergence
we may write
\vspace{-0.15cm}   
\begin{equation}
\label{eq:gmexy}
 \ql(\x,\y) = \intxy \!\! \dxx \!\! \dyy \, \qi(\xx,\yy) \, r\xyxy 
\vspace{-0.3em}
\end{equation}
where the kernel $\,r\,$ satisfies the
following three conditions:  
\vspace{-0.5cm}   
\begin{flalign}
\label{eq:Rintconstraint-pvsubstxy}
&&
\intxy \! \dx \! \dy \, r\xyxy
=
1 
~~ \forall \; (\xx\!,\yy) \! \,\in \Domain
\,,
\\
\label{eq:Rintconstraint-wt-avgxy}
&&
\intxy \!\! \dxx \!\! \dyy  \, r\xyxy
=
1 
~~ \forall \; (\x,\y) \! \,\in \Domain
\,,
\\
\label{eq:Rge0xy}
&\mbox{and}
\phantom{^{\big |}}  
&
r\xyxy
\;\geqslant\;   
0
~~
\forall \:
(\x,\y),\,(\xx\!,\yy) \! \,\in \Domain
\,,
\end{flalign}
but is otherwise arbitrary.   
Here  
we call $r\xyxy$ the
``\rf''
defining the \gm\ 
that takes place between the initial 
time and the later time.
The condition~(\ref{eq:Rintconstraint-pvsubstxy})  
ensures that
$r\xyxy$ represents a conservative redistribution of
PV substance  
in the sense that (\ref{eq:totalpv}) is satisfied.  
To  
show  
this, integrate (\ref{eq:gmexy})
with respect to $\x$ and $\y$ and then use
(\ref{eq:Rintconstraint-pvsubstxy})  
to deduce
(\ref{eq:totalpv})  
with $\delq = \ql - \qi$.  
The conditions~(\ref{eq:Rintconstraint-wt-avgxy})
and 
(\ref{eq:Rge0xy})
ensure that, for given $(\x,\y)$, \ 
$\ql(\x,\y)$ is a 
weighted average (with positive
or zero  
weights) of the initial PV values
$\qi(\xx\!,\yy)$.
This in turn ensures that
\gm\ cannot increase the range of
PV values, and
in particular
that an initially uniform PV profile
remains uniform.

We may think of  
$r\xyxy\dx\dy\sliver\dxx\Antisliver\dyy\sliver$
as the proportion of fluid transferred
from area $\dxx\Antisliver\dyy$ at location $(\xx,\yy)$
  to area $\dx\antisliver\dy$   at location $(\x,\y)$. \
Here ``fluid" has to be understood in a particular way.
The notional fluid, or material, has to be the sole transporter
of $q$-substance, whether by advection or by diffusion or otherwise.
That is, we imagine that different amounts of $q$-substance are
attached permanently to each fluid particle, so that, in particular,
the diffusivity of $q$ is the same as the self-diffusivity of the
notional fluid.
The notional fluid  
is incompressible,
as required by  
(\ref{eq:totalpv}),
(\ref{eq:Rintconstraint-pvsubstxy}),
and the concept of self-diffusivity.

The mathematical properties of the Green's function operators
  are further discussed in \citet{Shnirelman:1993}. 
  For instance, they form a partially-ordered semigroup.  The
partial ordering corresponds to successive mixing events.

For the PV-mixing problem we are mainly interested in
a zonally symmetric domain  
$|\yy|<L$;
and
sections~\ref{sec:gm}--\ref{sec:furtherimplications} 
will  
consider only zonally symmetric initial PV profiles, 
$\qi(\yy)$.
The PV distribution after \gm\ 
may or
may not be zonally symmetric. 
However, the \ahm\ change  
$\delM$ depends only on
$\qi(\yy)$ and on
the zonal or $x$ average  
of $\ql(\x,\y)$, 
denoted $\qlbar(\y)$.  
It is convenient to define
\begin{equation}
\label{eq:bigrdef}
\Ryy
:= 
  \int\!\! \dxx \, \overline{r\xyxy}
~,  
\end{equation}
the overbar again denoting the 
average with respect to $\x$ (not $\xx$).  
The zonal average of
(\ref{eq:gmexy})
is then  
\begin{flalign}
\label{eq:gme}
&& \qlbar(\y) &= \intd \dyy \, \qi(\yy) \Ryy ~, & 
\\
\label{eq:Rintconstraint-pvsubst}
&\mbox{where} &
\intd  \dy \, \Ryy  &= 1 \quad \forall \; \yy \! \in \D
~,
\\
&&
\label{eq:Rintconstraint-wt-avg}
 \intd  \dyy \, \Ryy  &= 1 \quad \forall \; \y \! \in \D 
~,
\\
&\mbox{and}  & 
\label{eq:Rge0}
\phantom{^{\big |}}  
 \Ryy &\geqslant 0 \quad \forall \; \y,\yy \! \in \D 
~,
\end{flalign}
(\ref{eq:Rintconstraint-pvsubst})--(\ref{eq:Rge0})
being the counterparts of  
(\ref{eq:Rintconstraint-pvsubstxy})--(\ref{eq:Rge0xy}).

A \rf\
$\R$ representing  
pure diffusion
is symmetric in the sense that 
$\Ryy$ $=$ $\R(\y,\yy)$.
This follows from the self-adjointness of the operator representing
the divergence of a downgradient diffusive flux. 
It is sometimes assumed that
all \rf s are symmetric,
but that would be   
too restrictive
for our purposes.  

\begin{figure*}
  \centering
  \noindent\includegraphics[width=33pc,angle=0]{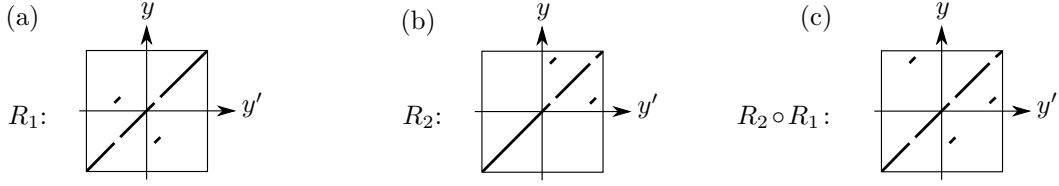}\\
\caption{\small  
Three \rf s of which the first two
are symmetric and the third asymmetric.
They represent  
purely advective rearrangements  
within the zone $|\y| < L$,
with no diffusive smearing.  
Values are zero except on the black  
sloping lines, which represent Dirac delta functions,
e.g. $\delta(\y-\yy)$ on the
main diagonal  
$\y = \yy$.
The first two  
\rf s
   $\Rs$ and $\Rf$   
describe simple exchanges of
small but finite  
(strip-like)  
fluid elements. 
The composite rearrangement
described by the third  
\rf\  
\mbox{$\RsRf$},
see (\ref{eq:composite}),  
is a cyclic
permutation among three
fluid elements, a
``three-cycle''
in group-theoretic terminology.
Notice
that the off-diagonal delta functions line up with
the gaps, or zeros, in the main diagonal.
They line up
both in the $\y$ direction
and in the $\yy$ direction so that both
(\ref{eq:Rintconstraint-pvsubst}) and
(\ref{eq:Rintconstraint-wt-avg}) are satisfied.
}
 \label{fig:Rcomposite}
\end{figure*}

Consider the examples of  
purely advective rearrangement  
in \fig \ref{fig:Rcomposite}.
The first two examples,
   with \rf s $\Rs(\yy,\y)$ and $\Rf(\yy,\y)$,  
are symmetric.  They  
correspond to patterns in
the $(\yy,\,\y)$ plane that are
mirror-symmetric about the
main diagonal,  
representing simple pairwise diffusionless
exchanges of fluid elements.
The third example 
depicts
the effect of
$\Rs$ followed by $\Rf$,
giving
\begin{equation}
 \qlbar(\y) 
= \intd \dyy \intd \dyyy \, \qi(\yyy) \Rs(\yyy,\yy) \Rf(\yy,\y) 
~.
\end{equation}
That is, 
the effect of
$\Rs$ followed by $\Rf$
is described by the
composite  
\rf
\begin{equation}
\label{eq:composite}
 \RsRf(\yyy,\y)
\;:=\,  
\intd \dyy \,  \Rs(\yyy,\yy) \Rf(\yy,\y)
~,
\end{equation}
which is asymmetric.
It represents a cyclic permutation
of three fluid elements  
and  
is the simplest kind of  
asymmetric \rf.
To be 
completely 
general we need to include such cases  
and their
elaborations.

In section \ref{sec:genproof}  
and Appendix~A 
we use the fact that
purely advective rearrangements are reversible,  
hence described by invertible mappings. 

\section{$\Mabs$ in terms of $\q$ for \sw}
\label{sec:Mdef}

For \shw\ beta-channel dynamics  
we may define $\Mabs$ as
the total \azm\ per unit  
zonal ($\x$) distance.  
Let the
\shw\ layer have
depth $H-\topogfull(\x,\y,t) +  
h(\x,\y,\timevar)$, \
where
$H$ is constant,  
$h$ is the free\break  
surface elevation,
$\topogfull$ is the bottom topography, 
and\break  
$h\ll H$, \ $\topogfull\ll H$. \
We assume $\topogmean = \topogmean(\y)$. \  
The fluctuating part
$
\topogfluct(\x,\y,t)  
:=  
\topogfull - \topogmean
$
can provide a 
quasi-topographic 
form stress 
to change $\Mabs$ and catalyze PV mixing,
as may happen in Jupiter's 
stratified 
weather layer.  
We choose the Coriolis parameter to be a constant,
$f_0$, thus regarding
the beta effect as due to the northward or
$\y$ gradient of the
zonally averaged bottom profile $\topogmean(\y)$,
corresponding to the latitudinal gradient of
Taylor--Proudman layer depth in the middle latitudes of
a spherical planet.   
Let    
$\dens$ be the constant mass density and
$u(\x,\y,\timevar)$ the zonal velocity with
$\overline{u}(\y,\timevar)$ its zonal average.
Then
to \qg\ accuracy  
\begin{align}
\Mabs 
&=
\dens \intd \! \dy \: 
\overline{\left(H + h -\topogfull \right)  
\left(u - f_{0} \y \right)}
\\
&= \dens H \intd \! \dy \:
  \left(
       \overline{u}
      - f_{0} \y \frac{{\overline{h}-\topogmean}}{H}
      - f_{0} \y 
  \right)
\\
&= \dens H \intd \! \dy \:
  \left(
       \overline{u}
      - f_{0} \y \frac{{\overline{h}-\topogmean}}{H}
  \right)
  +
  \mbox{const.}
\end{align}
Introducing the \qg\ stream function
$\psi = g h / f_{0}$ and
the Rossby deformation length
$\LD =  \sqrt{g H}/f_{0}$ we have
\begin{align}
\label{eq:M-psi}
\Mabs &=
\dens H \intd \! \dy \:
  \left(
      - \frac{\partial\overline{\psi}}{\partial\y}
      - f_{0} \y \frac{\overline{h}-\topogmean}{H}
  \right)
  +
  \mbox{const.}
\\
\label{eq:M-psi-intd}
&=
 \dens H
      \intd \! \dy 
        \left(
          \frac{\partial^{2} \overline{\psi}}{\partial \y^{2}}
          - \LD^{-2} \overline{\psi}
          + \beta\y
        \right)
        \y
  +
  \mbox{const.}\,,   
\end{align}
where the first term has been integrated by parts.  
We have
defined
\begin{equation}
\label{eq:beta-def}  
\beta\y
:=
   \frac{f_0
	 \topogmean}{H}  
\end{equation}
and assumed that
the Phillips boundary condition holds, 
namely
\begin{equation}
\label{eq:phillips}
\frac{\partial\sliver\overline{u}}{\partial t}  
\,=\,
-
\frac{\partial^2\overline{\psi}}{\partial\y\Sliver\partial\timevar}
\,=\,
0
\quad
\mbox{on $\y = \pm L$}
~,
\end{equation}
implying that   
the boundary term
\begin{equation}
\label{eq:ex-phillips-bc}  
-
\dens H
\left[
    \frac{\partial\overline{\psi}}{\partial\y}\y
\right]^{+ \! L}_{-\!L}
=
\mbox{const.}
\end{equation}
The Phillips boundary condition is
the standard way of stopping
mass and \am\ from leaking across
the side boundaries
\citep{Phillips:1954}.
Denoting the variable part  
of $\Mabs$
in (\ref{eq:M-psi-intd})  
by $\Mvar$
and defining $\q$ in the standard way,
ignoring a contribution $f_0$, 
as  
\begin{equation}
\label{eq:qdef}
\q
:=  
\nabla^2\psi - \LD^{-2} \psi + \beta\y
~,
\end{equation}
we have  
\begin{equation}
\label{eq:kelvin}
\Mvar
=
\dens H
\intd \! \dy \, \overline{\q}(\y) \, \y  
~.
\end{equation}
This expression has an
alternative
interpretation as the \textit{Kelvin impulse} for the \qg\ system,
per unit zonal distance
\citep[e.g.,][]{Buhler:2009}.  
Initially
\begin{equation}
\label{eq:Mhati}
\Mvar
=
\Mvari
:=  
\densintd \!\! \dy \;\, \qi(\y) \, \y ~.
\end{equation}
At the later  
time after \gm, the averaged $\q$ becomes
$\qlbar = \qi + \Delta \overline{\q}$,
so that
\begin{flalign}
\label{eq:Mhatl}
\Mvar
=
\Mvarl
&:=  
\densintd \!\! \dy \;\, \qlbar(\y) \, \y 
~,
\\
        &= \densintd \!\!\dy \intd \!\!\dyy \; \qi(\yy) \Ryy \, \y
~,
\\
  &=\Mvari + \delM
\end{flalign}
say, with 
\begin{flalign}
\label{eq:DMhat1}
 \delM 
&= \densintd \! \dy \; \Delta \qbar(\y) \, \y 
~,
\\
\label{eq:DMhat1b}
&= \densintd \! \dy \intd \!\! \dyy \; \qi(\yy) \Rdyy \, \y 
~,
\end{flalign}
where 
\begin{equation}
\label{eq:Rtilde}
\Rdyy
:=    
\Ryy - \delta(\yy-\y)
~,
\end{equation}
the difference between  
the \rf\ $\Ryy$  and  
the do-nothing \rf\ 
$\delta(\yy -\y)$.  
Here $\delta$ denotes the Dirac delta function.

\section{$\Mabs$ in terms of $\q$ for other systems}
\label{sec:other-systems}

The relations in 
section~\ref{sec:Mdef}
extend
straightforwardly to 
the sphere and
to a stratified
\qg\ system in a beta-channel.

In the stratified system,
with say a bottom boundary at 
pressure altitude  
$z = z_{0}$,
the PV is redistributed
separately on each 
$z$ surface,  
and the buoyancy acceleration
$f_{0} \partial \psi / \partial z$
is redistributed on 
$z=z_{0}$.
Therefore, 
each altitude $z$ has its
own $\R$ and $\Rd$
functions,
$\R(\yy,\y\Sliver;z)$         
and $\Rd(\yy,\y\Sliver;z)$    
say.  To obtain a concise 
formulation we may define
the PV to 
include a 
delta function at $z = z_0$
following \cite{Bretherton:1966c},
\begin{align}
        \nonumber
\Q(\x,\y\sliver;z) :&=   
\nabla^2\psi 
+\frac{1}{\rho_{0}(z)}
\frac{\partial}{\partial z}
\left(\rho_{0}(z)\frac{f_{0}^{2}}{N(z)^{2}}  
        \frac{\partial \psi}{\partial z}
\right)+
\\
&+
 \frac{f_{0}^{2}}{N(z_0)^{2}}
 \frac{\partial \psi}{\partial z}\Sliver\delta(z-z_{0})
+ \beta\y 
  \;,  
\label{eq:Qdef}
\end{align}
where 
$\rho_0(z)$ is 
the background density, 
$N(z)$ is  
the background buoyancy frequency, and
$\nabla^2$ still denotes the horizontal Laplacian.
If there is  
a rigid top boundary,  
then a  
further  
delta function can be added.
The initial and later $\Mvar$ values  
and the difference between them
are now, respectively,
\begin{align}   
 \Mvari &= \int \!\! dz \, \rho_{0}(z)\Antisliver 
        \intd \!\! \dy \;\, \Qi(\y\Sliver;z)\,\y
~, 
\\
 \Mvarl &= \int \!\! dz \, \rho_{0}(z)\Antisliver 
        \intd \!\! \dy \intd \!\!\dyy \; \Qi(\yy;z) \R(\yy,\y\Sliver;z)\,\y
~,
\end{align}
and
\begin{align}
\label{eq:DelM-layerwise}
\Delta\Mvar &= \int \!\! dz \, \rho_{0}(z)\Antisliver 
        \intd \!\! \dy \intd \!\!\dyy \; \Qi(\yy;z) \Rd(\yy,\y\Sliver;z)\,\y
~.
\end{align}
The contributions to $\Mvar$ add up layerwise
because PV inversion is a linear operation in \qg\ dynamics. 

These relations also extend to
a single layer on a sphere, provided that $\LD=\infty$ 
and that
\azm\   
per unit zonal distance
is replaced by \aam\
per radian of longitude.
Then
the counterpart of
(\ref{eq:DMhat1}) 
is
\begin{equation}
 \label{eq:Msphere}
\delM 
=
\rho_{0}H a^{4} \!            
\int^{1}_{-1} \!\! d \mu \,
\Delta \overline{q}(\mu)\,\mu  
~,
\end{equation}
where 
$a$ is the radius of the sphere, 
$\mu := \sin \phi $  
where
$\phi$ is the latitude,
and $q$ now  
denotes the absolute vorticity.
The \ghm\ conditions 
(\ref{eq:gme})--(\ref{eq:Rge0})
and
the formulae
(\ref{eq:Mhati})--(\ref{eq:Rtilde})
apply to 
the  
sphere  
provided that
$\y$ is replaced by $\mu$,
$\rho_0 H$ by $\rho_0 Ha^{4}$,  
and  $\pm L$ by $\pm 1$.

\section{The basic theorem}
\label{sec:basictheorem}

In this section we  
prove the basic theorem  
that $\delM$ is always negative
for monotonically increasing
$\qi(\yy)$ 
and
any nontrivial  
rearrangement function $\R$
such that
integrals like (\ref{eq:DMhat1b})
make mathematical  
sense,  
with values independent of the
order of integration.
The same proof will apply to the
potential-energy problem, 
with zonal averaging replaced by horizontal area integration 
for general container shapes,
as explained in section~\ref{sec:ape}.

Nontrivial means
``do something'' rather than  
``do nothing'':
$\Rd$
in (\ref{eq:Rtilde})  
must be nonvanishing in an appropriate sense. 
More precisely,
nontrivial means
that  $\Ryy$ and $\Rdyy$
have nonvanishing off-diagonal values somewhere, 
where those off-diagonal values 
have \emph{nonzero measure}
in the sense that they can make nonzero
contributions to integrals like
(\ref{eq:DMhat1b}).
This in turn means that the nonvanishing off-diagonal values
must exist in some finite neighborhood, albeit possibly
a neighborhood in the form of a line segment,  
as in the delta-function examples of
\fig\ref{fig:Rcomposite}.

\Eq(\ref{eq:DMhat1b}) can 
be rewritten
\begin{align}
 \label{eq:DMhat2}
\delM &= \densintd \dyy \; \qi(\yy) \Sliver \eta(\yy)
~,
\end{align}
where by definition
\begin{align}
\label{eq:eta}
\eta(\yy) &:= \intd \! \dy \; \Rdyy \, \y  
~. 
\end{align}
In virtue of
(\ref{eq:Rtilde}), 
$\eta(\yy)$ may be regarded as the
average latitudinal displacement of fluid initially at $\yy$.
Denote the indefinite integral of $\eta(\yy)$
by  
$\Etabig(\yy)$ (cyrillic-style  
big                   
Eta).                
Specifically,  
\vspace{-0.2cm}
\begin{align}
\label{eq:Exch1}  
\Etabig(\yy)
&\,:=   
 \int_{-L}^{\yy} \!\!\!\! \dyyy \, \eta(\yyy) 
\,=\!           
    \int_{-L}^{\yy} \!\!\!\! \dyyy \!\!
\intd \! \dy   \, \Rd \left(\yyy,\y\right)  \y
\\
\label{eq:Exch2}
&\;=\;
  \int_{-L}^{\yy} \!\!\!\! \dyyy \!\!
\intd \! \dy   \, \Rd(\yyy,\y)
   \left(
        \y-\yy
   \right)
\\
\label{eq:Exch3}
&\;=\;
 -
  \int_{\yy}^{L} \!\!\!\! \dyyy \!\!
\intd \! \dy   \, \Rd(\yyy,\y)
   \left(
        \y-\yy
   \right)
~,
\end{align}
where the penultimate step  
uses  
(\ref{eq:Rintconstraint-pvsubst})
and
(\ref{eq:Rtilde}),
implying that 
$\textstyle\intd\dy\Sliver\Rd(\yyy,\y)=0$ for all $\yyy$,
and the last step
(\ref{eq:Rintconstraint-wt-avg}) 
and
(\ref{eq:Rtilde}),
implying that
$\textstyle\intd\dyyy\Sliver\Rd(\yyy,\y)=0$ for all $\y$.
The last step depends on interchangeability of the order of
integration.  
From (\ref{eq:Exch2}) and (\ref{eq:Exch3})
we see
that
$\Etabig(-L)=0=  
 \Etabig(+L)$.  
Therefore
(\ref{eq:DMhat2})
may be integrated by parts 
to give
\vspace{-0.3cm}
\begin{equation}
\label{eq:DMhat3}
\framebox{$
\phantom{\Bigg]^I}  
 \delM = - 
\densintd \dyy \,
\frac{\partial \qi(\yy)}{\partial \yy}\, 
\Etabig(\yy)  
~.
  \quad
$}
\end{equation}
So if, finally,  
for nontrivial $\R$,
we can
prove that $\Etabig(\yy)$ 
is nonnegative
for all values of $\yy$
and 
nonvanishing with nonzero measure for at least some  
values
of $\yy$, then
the theorem will follow.  That is,
(\ref{eq:DMhat3})
will then imply that
\begin{empheq}[box=\framebox]{align} 
\phantom{\Big|}  
\label{eq:basictheorem}
\delM<0
\;~\mbox{ if }~\:  
{\partial \qi(\yy)}/{\partial \yy}>0
~\mbox{ for all $\yy$~,}
\phantom{\Big|}  
\end{empheq} 
and vice versa.
That is,
{the sign of $\delM$ must always be opposite to
the sign of the initial monotonic  
PV gradient.}

\smallskip  

To prove that $\Etabig(\yy)$
is nonnegative
we rewrite (\ref{eq:Exch3}),
after changing the order of integration,  
as
\vspace{-0.2cm}
\begin{align}
        \nonumber
\Etabig(\yy)
&\,=
\int_{-L}^{\yy} \!\!\!\!\!\dy\!
\int_{\yy}^{L}  \!\!\!\!\!\dyyy
      \,\Rd(\yyy,\y)
      \left|
           \y-\yy
      \right|
      \\
\label{eq:Exch4}
\;&-
\int_{\yy}^{L} \!\!\!\!\!\dy\!
\int_{\yy}^{L} \!\!\!\!\!\dyyy
      \,\Rd(\yyy,\y)
      \left|
           \y-\yy
      \right|
\;.
\end{align}
Again because
$\textstyle\intd\dyyy\Sliver\Rd(\yyy,\y)=0$ for all $\y$,
we may replace
$\textstyle\int\!\!_{\yy}^{L}\dyyy$\,
by
$\,-\!\!\textstyle\int\!\!_{-L}^{\;\yy}\dyyy$.
Applying this to the second term only, we obtain an
expression in which
$\Rd$ can be replaced by the nonnegative
function $\R$,
\begin{align}
        \nonumber
\Etabig(\yy)
&\,=
\int_{-L}^{\yy} \!\!\!\!\!\dy\!
\int_{\yy}^{L}  \!\!\!\!\!\dyyy
      \,\R(\yyy,\y)
      \left|
           \y-\yy
      \right|
      \\
\label{eq:Exch5}
\;&+
\int_{\yy}^{L}  \!\!\!\!\!\dy\!
\int_{-L}^{\yy} \!\!\!\!\!\dyyy
      \,\R(\yyy,\y)
      \left|
           \y-\yy
      \right|
\;,
\end{align}  
because
there are no contributions from
the main diagonal
$\y = \yyy$. \
For given $\yy$,
the two rectangular domains of integration
for (\ref{eq:Exch5})
intersect each other and the main diagonal at a single point
only,
$\y = \yyy = \yy$. \
(The two domains  
are mirror images of each other in the main
diagonal.) \
At the point $\y = \yyy = \yy$ the factor
$\left|\y-\yy\right|$
is zero, annihilating any delta functions.
Therefore $\Etabig(\yy)$
is nonnegative.  

Now  
as $\yy$ runs from $-L$ to $L$,
the two
domains  
sweep over the upper and lower triangles of the square
$-L \leqslant \y   \leqslant L$, \
$-L \leqslant \yyy \leqslant L$, \
together  
covering the entire square.
By definition,
a nontrivial \mbox{$\R$ function} must have nonzero measure
somewhere off the main diagonal,
in some finite neighborhood of  
a location
with $|\y|\ne L$,
$|\yy|\ne L$,
and
$\y-\yy \ne 0$.  
Whichever moving domain encounters that location must
continue to intersect it as $\yy$ runs through some
finite range of values,
implying that $\Etabig(\yy)>0$
over that finite range.  So $\Etabig(\yy)$
is not only nonnegative, but also
nonvanishing with nonzero measure, for
any  
nontrivial $\R$,  
and the theorem follows.

An alternative proof using an entirely different
nonnegative function is given in 
Appendix~A. 

\section{Connection to available potential energy}
\label{sec:ape}

The basic theorem can 
alternatively 
be read as governing the sign of
the potential-energy change due to 
three-dimensional   
generalized mixing
of a Boussinesq fluid  
within a fixed container  
in a uniform gravitational field.  

Consider first 
a container with vertical walls.  Then 
(\ref{eq:DMhat3})
carries over at once  
if we read
$\qi$ as the buoyancy
acceleration, $\yy$  
as the altitude,
the $\R$ and $\Rd$ functions as applying to
horizontal area averages,
starting from a three-dimensional version of the
$r$ function in
(\ref{eq:gmexy})--(\ref{eq:Rge0xy}),  
and
$\delM$ as proportional to 
minus  
the potential-energy change.
Second,  
consider a container of arbitrary shape $\cal V$
as being embedded
within the vertical-walled
container.
We merely extend the definition of $r$
and hence of $\R$ and $\Rd$ such that
no \gm\  
takes place outside $\cal V$. \
With this understanding
(\ref{eq:DMhat3}) still applies, and 
(\ref{eq:basictheorem}) follows.
That is, if the initial state
is undisturbed and stably stratified,
with the same stratification at
all horizontal positions (including those in any separate
``abyssal basins''), then the potential-energy
change is guaranteed to be positive
for any nontrivial $\R$ whatever.   
This generalizes a standard result
in the theory of
available potential energy 
saying the
same thing for a purely advective 
$\R$ \citep[e.g.,][Appendix~A below]{Holliday:1981}. 

We emphasize that the generalized result
depends on having a linear equation of state, 
as is standard  
for  
Boussinesq models, 
since
only then is the buoyancy acceleration
a transportable, mixable
quantity.\footnote{For
more general equations of state, especially those containing
thermobaric terms,
there is no straightforward concept of potential energy.
As first shown by W. R. Young, the Boussinesq limit then
needs reconsideration, and the consequences are
nontrivial.
It turns out that potential energy has to be replaced by
a ``dynamic enthalpy'' that contains both gravitational
and vestigial thermodynamic contributions 
\citep{Young:2010}. 
Such generalized
Boussinesq models are outside  
our scope here.  
}

\section{The physical meaning of $\Etabig(\yy)$}
\label{sec:Eta-meaning} 

Reverting to the PV interpretation,
with $\y$ northward rather than upward,
we consider the function
$\Sliver\Etabig(\yy)/(L+\yy)$. \
The definition (\ref{eq:Exch1}) shows that 
$\Etabig(\yy)/(L+\yy)$ is the average  
northward displacement  
of all the notional fluid  
initially south of $\yy$. 
Equivalently,
$\Etabig(\yy)/(L+\yy)$ is the
northward displacement  
of that
fluid's centroid.  This makes  
the nonnegativeness  
of $\Etabig$ more intuitively apparent.
The centroid is initially as far south as it can be,
and can therefore only move northward.
We may reasonably call
$\Etabig(\yy)$ itself the ``area-weighted bulk displacement'' of
all the fluid initially south of $\yy$, or
``\bdf'' for brevity.

The fact that $\Etabig(L)=0$ expresses  
what can also, now, be seen to be intuitively reasonable,
namely  
that there can be no bulk displacement of the entire
zone $-L\leqslant\yy\leqslant L$.
The  
fluid has nowhere to go.  Its centroid must
remain fixed under any generalized mixing operation confined
to the
zone $-L\leqslant\yy\leqslant L$.
And the symmetry expressed by
(\ref{eq:Exch5})  
says that we may equally well think of
$\Etabig(\yy)$
as the \emph{southward}
area-weighted bulk displacement of all the fluid initially
\emph{north} of $\y=\yy$.

\begin{figure}[t]
   \centering
   \noindent\includegraphics[width=10pc,angle=0]{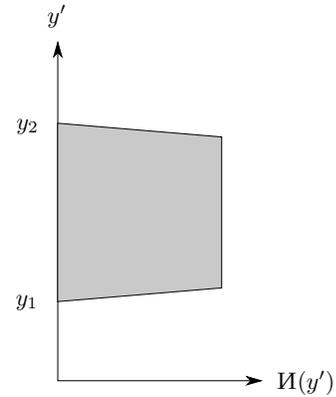}\\
\caption{\small  
    Illustration of 
    $\Etabig(\yy)$
    for a
    \rf\ that simply exchanges material  
    between latitudes
    $\y_{1}$ and $\y_{2}$, as for instance in
    \fig\ref{fig:Rcomposite}b.   
The finite slopes  
near  
$\y_{1}$ and $\y_{2}$
are due to the finite widths of the fluid elements exchanged.
The maximum value of $\Etabig$ is $\y_2 - \y_1$.
  }
   \label{fig:Exchange}
\end{figure}

\Fig \ref{fig:Exchange}
shows
a simple example, the \bdf\ $\Etabig(\yy)$
corresponding to the \mbox{$\R$ function} shown in
\fig \ref{fig:Rcomposite}b. 
Nothing happens to the fluid south of
$\y_1$ and north of $\y_2$.  
However, 
there is, 
for instance, 
a northward bulk displacement of the fluid originally in 
$(-L, \yy)$ whenever $\yy$ lies between
$\y_1$ and $\y_2$.
The transitions across
$\y_1$ and $\y_2$
have small but finite widths, corresponding to the small but
finite
line segments  
in the off-diagonal regions of
\fig \ref{fig:Rcomposite}b.

The foregoing applies of course
to the potential-energy interpretation,
with northward and southward
replaced by 
upward and downward.

\section{Further implications, 
including generalized shear-instability theorems  
}
\label{sec:furtherimplications}

The   
result
(\ref{eq:DMhat3})
carries over to DS08's
case of a sphere with $\LD=\infty$,
with $\y$ replaced by $\mu$, the
sine of the latitude,  
and
$\delM$
replaced by its spherical counterpart,
the \aam\ increment 
(\ref{eq:Msphere}),
as noted
at the end of  
section~\ref{sec:other-systems}.
And   
(\ref{eq:DMhat3})
also  
carries over
to the stratified systems of
section~\ref{sec:other-systems},
with the factor $\rho_0 H$
replaced by a vertical integration
and $\q$ by $\Q$  
as in  
(\ref{eq:Qdef})--(\ref{eq:DelM-layerwise}).  
Therefore, the basic theorem (\ref{eq:basictheorem})
holds
in any case for which there are
monotonic profiles  
of $\Q$ on each 
of the levels  
subject to  
mixing, provided that
all the gradients
$\partial\Q/\partial y$   
have the same sign including
the gradients  
of the Bretherton delta function or functions.

It is worth noting the implications of such cases  
for the theory of \qg\ shear instability,
in particular the theorems of 
\citet{Charney:1962}
and \citet{Arnold:1965}.
These theorems in their original forms
apply only to
nondiffusive Hamiltonian dynamics,
and therefore only to
purely advective rearrangements.
The
basic theorem
(\ref{eq:basictheorem})
generalizes the   
Charney--Stern theorem and a case of
Arnold's first stability theorem  
--- which we call
``Arnold's zeroth stability theorem'',
or ``the Arnol'd theorem'' for brevity   
--- to
cover finite-ampli\-tude disturbances with
arbitrary amounts of PV mixing.
The Arnol'd theorem in question
is the nonlinear counterpart of
the Rayleigh--Kuo theorem, rather than the Fj\o rtoft theorem  
of which Rayleigh--Kuo is a special case.

In instability problems there are no
external sources or sinks of \aam.
Growing instabilities 
exchange angular momentum 
purely internally,   
through radiation or diffraction stresses.
This is possible,
the basic theorem tells us,
only if there are 
regions in which the 
$\q$ or $\Q$ gradients  
have different signs.
Conversely, whenever the 
$\q$ or $\Q$ gradients 
are
nonzero 
and all of one sign,
instability is impossible.
These are exactly the circumstances in which
the Charney--Stern theorem and the Arnol'd theorem  
were originally proved
for purely advective rearrangements.  
and can now be proved,
using (\ref{eq:basictheorem}), for
the  
far more general
redistributions
defined in section~\ref{sec:gm},  
which include PV mixing.

The proof runs as follows.
We
start 
with $\q=\qi(y)$,
or  $\Q=\Qi(y)$ on each level.  
An initial
finite-amplitude disturbance is set up advectively, by
undulating the PV contours. 
To do so requires artificial forcing.
This is because of the hypothesis  
that
the $\q$ or $\Q$ gradients are nonzero and 
all of one sign.
By (\ref{eq:basictheorem}),
$\Mabs$ must change by some nonvanishing amount $\delM$
during the setup.

We
then let the system run freely.
The  
free  
dynamical evolution may include
wave breaking and PV mixing ---
going beyond Hamiltonian evolution.  
PV invertibility implies that the free evolution  
can be fully described by specifying a succession of  
PV distributions.  Equivalently, therefore,
the free evolution  
can be described by a succession of \mbox{$\R$ function}s
operating on $\q=\qi(y)$ or   $\Q=\Qi(y)$. \
Each such function is the  
composite of two \mbox{$\R$ function}s,
the purely 
advective \mbox{$\R$ function} describing
the initial setup  
and one of the  
general \mbox{$\R$ function}s describing the subsequent
free evolution.  

The free evolution keeps $\delM$ constant.  
Since
   (\ref{eq:DMhat3}) or its $\Q$ counterpart,  
vertically integrated as necessary,  
is sign-definite by hypothesis,  
then either it or its negative
qualifies as a Lyapunov function
(from \mbox{$\R$ function}s to nonnegative real numbers), whose constancy
under 
free evolution  
implies neutral nonlinear stability.
This is the generalized Arnol'd's zeroth theorem.

We may remark that
the sign-definite function
(\ref{eq:DMhat3-casimir})
below also qualifies as a Lyapunov function,
vertically integrated as necessary,
providing an alternative proof.

\section{Nonmonotonic, zonally asymmetric $\qi$} 

\label{sec:genproof}

The basic  
theorem
(\ref{eq:basictheorem})
applies to zonally symmetric
and monotonic 
$\qi(\yy)$ only.
This is the most important
case, 
but it may be of interest to note
what can be proved for more general
initial conditions 
$\qi(\xx,\yy)$.

Consider a pair of 
PV distributions
$\qfxy$,  
$\qsxy$
that can be 
derived 
from each other by purely  
advective,
and therefore reversible,
rearrangement.
That is,
\begin{flalign}
\label{eq:rfwd}
&&
\qs(\x,\y) &= \intxy \dxx\Antisliver\dyy \;
\qfxy\,\rfwd(\xx,\yy,\x,\y)
\\
\label{eq:rback}
\raisebox{10pt}{and}  
&&
\qf(\x,\y) &= \intxy \dxx\Antisliver\dyy \;
\qsxy\,\rfwd(\x,\y,\xx,\yy) 
~,
&
\end{flalign}
where the \rf\ \,$\rfwd$\, describes
an invertible 
mapping.

For given $\rfwd$,
consider
the set of
all possible
\rf s $r$
together with
the set of all possible
composites $r \circ \rfwd$.
Because of reversibility,
the set of all $r$
must be the same
as the set of all $r \circ \rfwd$.
Therefore the set of all possible
$\Mvarl$ values 
that can result from applying
the $r$'s to 
an initial PV distribution $\qf$
must be the same as
the set of all possible
$\Mvarl$ values 
from applying
the $r$'s to an initial $\qs$.

For a general 
initial  
$\qf(\xx,\yy)$
we can always find an
advective rearrangement  
$\rfwd$ such that
$\qm$
is a monotonically increasing function of $\y$ alone
(Appendix B). 
Denote that function by
$\qm \left[ \y;\qf\functional \right]$.
The corresponding $\Mvar$ value
is 
\begin{equation}
\Mvar_{2} \left[ \qf\functional \right] 
= \intd \dyy \, \qm \left[ \yy;\qf\functional \right] \yy 
~.
\end{equation}
The basic theorem of
section~\ref{sec:basictheorem}
restricts the possible $\Mvarl$ values
that can be attained starting from
$\qm \left[ \y;\qf\functional \right]$.
Specifically,
\begin{equation}
\label{eq:genresult1}
\Mvarl
\;\leqslant\;
\Mvar_{2}  \left[ \qf\functional \right]
\sliver.  
\end{equation}
The same argument applies to the
monotonically decreasing
case.
Because the $\y$
origin is 
in the center of the
$\y$ domain,
the resulting $\qm$
function is simply   
$\qm \left[-\y;\qf\functional \right]$
and the corresponding $\Mvar$ value
is $-\Mvar_{2} \left[ \qf\functional \right]$.
In summary,
identifying $\qf(\xx,\yy)$
with our general initial condition $\qi(\xx,\yy)$,
we now have        
\begin{equation}
\label{eq:genresult2}
 -\Mvar_{2} \left[ \qi\functional \right]
\;\leqslant\;
\Mvarl
\;\leqslant\;
\Mvar_{2}  \left[ \qi\functional \right]
\Sliver.  
\end{equation}
That is,
the two possible extreme
values  
of $\Mvarl$
correspond to the two extreme,
monotonically decreasing or increasing,
zonally symmetric profiles into which  
$\qi(\xx,\yy)$ can be advectively  
rearranged.

\begin{figure}[t]  
\centering
  \noindent\includegraphics{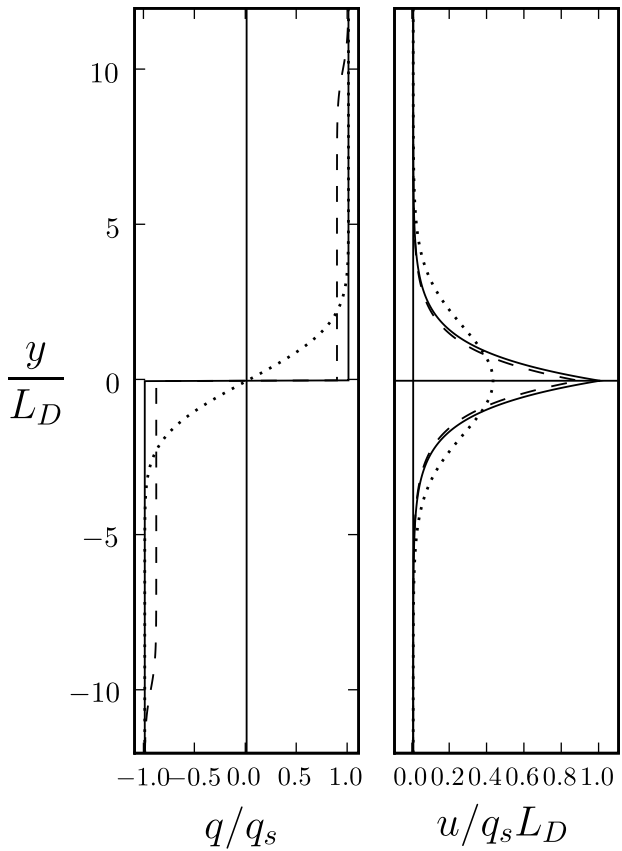}\\
\caption{\small  
The \jhrs\ thought experiment.
An initial PV profile in the form of a
step function (solid curves)
is smeared diffusively 
(dotted curves).
This smeared profile is then 
resharpened by mixing PV on the flanks
of the jet
(dashed curves).
Here,
as throughout this paper,
``mixing'' entails  
conservation of PV 
substance (\ref{eq:totalpv}).
}
\label{fig:resharp1}
\end{figure}

\section{The simplest \jhrs\ problem}  
\label{sec:resharp}

Consider the following \shw\ thought experiment
in an unbounded domain, $L=\infty$.  
We begin with a perfectly 
sharp jet,
with concentrated PV gradients at its core
(solid curves in
\fig \ref{fig:resharp1}). 
First, the concentrated PV gradients  
are smeared out,
decelerating the jet
and decreasing
the \aam\  
$M$
(dotted curves in
\fig \ref{fig:resharp1}). 
Second, the PV is mixed
on both sides  
of the jet, 
resharpening and
accelerating it
(dashed curves in
\fig \ref{fig:resharp1}). 
Perhaps
counterintuitively,
the basic~theorem (\ref{eq:basictheorem})  
implies that
$M$ must decrease further,
at this second stage.
even though the jet core accelerates.  
Let us look at what happens
in more detail.

Consider 
the \qg\ \shw\ system
with the initial PV profile
in the form of a step
of size $2 \bigjump$,
\begin{equation}
\qi(\yy) = 
\left\{
  \begin{array}{rl}
     \bigjump  & \mbox{ $(\yy > 0)$} \\
    -\bigjump  & \mbox{ $(\yy < 0)$}  
~.
  \end{array}
\right.
\end{equation}
Inversion gives the familiar  
velocity profile 
\begin{equation}
 u_{i}(y')=
 \bigjump\sliver \LD \exp\left(-\frac{\left|y'\right|}{\LD}\right) 
~,
\end{equation}
shown by the solid curve in
\fig \ref{fig:resharp1}b. 
After the first stage,
in which the concentrated gradients are smeared out,
the PV profile is taken 
in error-function form
\begin{equation}
\label{eq:qsmear}
\q_1(\y)
\,=\,
\frac{2\bigjump}{L\sqrt \pi} 
\int^{\y}_{0} \!\!
d\ydumm \,
   \exp\!
   \left(
       -\Sliver\frac{\ydumm^{2}}{L^2}
   \right)
~,
\end{equation}
shown by the dotted curve in \fig \ref{fig:resharp1}a,
for which the lengthscale $L$ has been taken as $2\LD$.
Inversion gives  
the corresponding 
smeared 
velocity profile,
shown by
the dotted curve in 
\fig
\ref{fig:resharp1}b,
as
\begin{equation}
\label{eq:usmear}
u_{1}(\y)
=
\frac{\LD\sliver\bigjump}{L\sqrt{\pi}}  
\int^{\infty}_{-\infty} \!\! d \ydumm \, 
\exp
\left(
  -\frac{\left|\y-\ydumm\right|}{\LD}
  -\frac{\ydumm^{2}}{L^{2}}
\right) 
\end{equation}
[cf.\ (\ref{eq:uinversion})ff.].
The change in $M$
due to the PV 
redistribution in 
this first stage
is 
\begin{equation}
\delM_{1} 
=
\int^{\infty}_{-\infty} \!\! d \y(\q_{1}-\qi)\,\y 
= 
-\frac{\bigjump L^{2}}{2}
~,
\end{equation}
as can be verified from an integration by parts.  

After the second stage,
the PV has been perfectly mixed 
on either side of the jet core (dashed curves) 
out to fringes at around 
$|\y|$ $=$ $\lambda \LD$, say,
where $\lambda \gg 1$.
In the figure, 
we have taken $\lambda$ $=$ $10$.
Within the two perfectly mixed regions,
the resharpened PV distribution  
is 
\begin{flalign}
&& \q_{2}(\yy) &= 
\left\{\!\!
  \begin{array}{ll}
     \quad\:\bigjump - \littlejump  & \mbox{ $(\yy > 0)$} \\
    -\left(\bigjump - \littlejump\right)   & \mbox{ $(\yy < 0)$} 
  \end{array}
\right.
&
\\
\nonumber
&\mbox{where}&&&
\\
\label{eq:littlejump}
&&\littlejump &= (\lambda \LD)^{-1} \left(\bigjump L/ \sqrt{\pi}\right)  \ll \bigjump 
~.&
\end{flalign}
This
assumes fringes antisymmetric about
$\y$ $=$ $\pm\lambda\LD$, 
as well as 
total PV conservation, 
\eq(\ref{eq:totalpv}),
and neglect of the 
Gaussian tails
in (\ref{eq:qsmear})   
for $\y$ $\gg$ $\LD$.
The corresponding
resharpened 
velocity profile is
\begin{equation}
 u_{2}(y')=\left(\bigjump - \littlejump\right)\sliver \LD \exp\left(-\left|y'\right|/\LD\right) 
~,
\end{equation}
provided that the peripheral 
fringes have length scales 
$\gg$ $\LD$. 
(Narrower peripheral fringes, 
not 
$\gg$ $\LD$,
would invert to give
two extra jets, albeit weak ones.)

The change in $M$ 
due to the PV redistribution in the second stage is
\begin{align}
        \nonumber
        \delM_{2} &= -\delM_{1} + \int^{\lambda \LD}_{-\lambda \LD}\!\! \dy \, (\q_2 - \qi)\, \y 
        \\
				&= -\bigjump L^{2} \left\{ \lambda \left( \frac{\LD}{L\sqrt{\pi}} \right) - \frac{1}{2} \right\}
~.
\end{align}
Because $\lambda\gg 1$,  
$\left|\delM_{2}\right|$ 
$\gg$ 
$\left|\delM_{1}\right|$.
In this example, $M$ not only decreases at each stage,
as the basic theorem says it must,
but the decrease is far greater at the second stage,
even though the jet core still accelerates.
The total change $\delM$ over both stages
is
\begin{equation}
 \delM = 
 \delM_{1} + \delM_{2} =
   -\bigjump L^{2}  \lambda 
   \left( 
       \frac{\LD}{L\sqrt{\pi}} 
   \right)
~.
\end{equation}
This can also be written
\begin{equation}
\delM =
-\frac{\bigjump L^{2}}{\pi}
   \left(
      \frac{\littlejump}{\bigjump}
   \right)^{-1}
~.
\end{equation}
As $\littlejump/\bigjump$ decreases,
the PV profile $\qs$ returns closer and closer to
the initial PV profile $\qi$
while
$\delM$ becomes increasingly large and negative.
There is an increasingly
large cost associated with mixing far from
$\y = 0$.
In the case of
\fig \ref{fig:resharp1},
$\littlejump/\bigjump \approx 0.11$. \ 
Furthermore
$\delM_{1}$ $=$ $-2 \bigjump \LD^{2}$
and 
$\delM_{2}$ $\approx$ $-9 \bigjump \LD^{2}$.
If 
$\littlejump/\bigjump$
were decreased
to $0.01$, then
$\delM_{2}$ would
become $\approx$ $-125 \bigjump \LD^{2}$.

\section{General jet sharpening}  
\label{sec:genjet}

Consider a more
general \shw\
thought experiment,
now starting from a 
general
monotonic
PV
profile $\qi(\yy)$.
For definiteness,
we take the monotonically
increasing case
${\partial \qi(\yy)}/{\partial \yy}>0$.

We suppose that \gm\ 
takes place except that
there 
is
no mixing
across 
a particular
material contour initially at
latitude $\y=\y_{0}$.
That is,
the contour
behaves as an  
eddy-transport barrier.  
The contour
may
undulate
during the mixing,
but we assume that
it straightens out afterwards
and returns to latitude $\y_{0}$\sliver,  
consistent with \qg, area-preserving advection.  
The net effect of 
the mixing 
can then be described by a
nontrivial  
zonally-averaged \rf\ $\Ryy=\Ryyy$,
recall (\ref{eq:bigrdef}),
such that
\begin{equation}
\label{eq:block-diagonal}  
\Ryyy = 0
\mbox{~~~if~~}
\y < \y_{0} < \yy
\mbox{~~~or~~}
\yy < \y_{0} < \y
\sliver.
\end{equation}
It will be proved that,
in this thought experiment,
for finite $\LD$,  
the net change $\Delta\uoverline(\y_{0})$
in the zonal-mean zonal flow
  at $\y=\y_0$  
is always positive.

In particular,  
we may choose 
the material contour
$\y=\y_0$ 
to be
in 
the core of a jet.
So
\emph{mixing on one or both flanks of a jet must always
accelerate the
straightened-out  
jet core,  
regardless of
the details of the mixing provided only that the jet core
has persisted, throughout,  
as an eddy-transport barrier.}  Mixing could be confined, for
instance, to locations arbitrarily far from the jet core,
though of course the resulting $\Delta\uoverline(\y_{0})$
would then be small.  

Differentiating 
the expression for $\q$
in (\ref{eq:qdef})
with respect to $\y$,
and taking the zonal average, we obtain
the inversion problem  
for the change
$\Delta \uoverline(\y)$ 
in $\uoverline(\y)$ due to an
arbitrary change 
$\Delta \bar \q(\y)$
in $\bar \q(\y)$,
\begin{align}
\label{eq:uinversion}
\left(
\frac{\partial^2}{\partial \y^2}
-
\LD^{-2}
\right)
\Delta \uoverline(\y)
=
-
\frac{\partial \Delta\bar\q(\y)}{\partial \y}
~.
\end{align}
This can 
be solved 
with the Green's function $\G(\y,\y_0)$
defined by
\begin{align}
\label{eq:greens}
\left(
\frac{\partial^2}{\partial \y^2}
-
\LD^{-2}
\right)
\G(\y,\y_0) 
&=
-
\delta(\y-\y_0)
\end{align}
with
$\G(\y,\y_0)$ vanishing on the boundaries $\y\pm L$  
to satisfy the Phillips boundary condition
(\ref{eq:phillips}).
The proof will apply both to finite and to infinite $L$
(though
$\LD$ has to be finite).  
We have  
\begin{align}
\label{eq:greens-soln}
\Delta\uoverline(\y_{0}) 
&=
\int_{-L}^{L} \!\! \dy \,
\G(\y,\y_0)
\frac{\partial}{\partial \y}
   \Delta\bar\q(\y)
~,
\end{align}
as can be verified by subtracting
$\G(\y,\y_0)$
times (\ref{eq:uinversion}) 
from
$\Delta\uoverline(\y)$ 
times (\ref{eq:greens})
and integrating with respect to $\y$. \
Taking  
   $\Delta\bar\q$ 
$=$ 
$\textstyle\intd\Sliver\dyy\sliver\qi(\yy)\sliver\Rdyyy$,  
    with $\Delta\R$    defined by (\ref{eq:Rtilde}),  
we may integrate (\ref{eq:greens-soln})  
by parts to give
\begin{align}
\label{eq:deltau2}
\Delta\uoverline(\y_{0}) &=
-\int^{L}_{-L} \!\! \dyy \, 
\qi(\yy)
\etamod (\yy,\y_0)
\end{align}
where  by definition 
\begin{align}
\label{eq:etamod}
\etamod (\yy,\y_0)
:=
\int^{L}_{-L} \!\! \dy \,
\Rdyyy\,
\newpartialmetric(\y,\y_0) 
\end{align}
with
$\newpartialmetric(\y,\y_0) := \partial\G(\y,\y_0)/\partial\y$. \
Now let
\begin{equation}
\Etamod(\yy\Antisliver,\sliver\y_0) :=
\int_{-L}^{\yy} \!\!\! \dyyy \,
   \etamod(\yyy\Antisliver,\sliver\y_0) 
~.
\end{equation}
The reasoning below
(\ref{eq:eta})--(\ref{eq:Exch3})  
applies word-for-word to the functions
$\etamod$ and $\Etamod$,
after replacing  
the right-hand factors $\y$ and $\y-\yy$ in
(\ref{eq:eta})--(\ref{eq:Exch3}) 
by $\newpartialmetric(\y,\y_0)$ and
$\newpartialmetric(\y,\y_0) \!-\! \newpartialmetric(\yy,\y_0)$
respectively, $\y_0$ being fixed throughout.  
It follows that
$\Etamod(-L,\y_0)$ $=0$ $=\Etamod(+L, \y_0)$. \
Integrating 
(\ref{eq:deltau2})
by parts,
we therefore get a result analogous to (\ref{eq:DMhat3}),
\begin{equation}
\label{eq:generalsharpen}
\framebox{$
\phantom{\Bigg]^I}  
 \Delta\uoverline(\y_{0}) =
\int^{L}_{-L}\!\! \dyy \,
\frac{\partial \qi(\yy)}{\partial \yy}\, 
\Etamod(\yy\Antisliver,\sliver\y_0)  
~.
  \quad
$}
\end{equation}
We now use
the eddy-transport-barrier assumption
(\ref{eq:block-diagonal}).
The assumption says that $\Ryyy$ and $\Rdyyy$ have  
a block diag\-onal structure
in the $\yy\antisliver\y\Sliver$ plane,
with nonvanishing values confined to  
two diagonal blocks meeting at
$\y=\yy=\y_0$. \
If $\yy > \y_0$, then nonvanishing contributions to
$\Etamod(\yy,\y_0)$
come from the upper right block only, and
if $\yy < \y_0$ from the lower left only.
Within each block
$\newpartialmetric(\y,\y_0)$
is a monotonically increasing function of $\y$,
as will be shown next,  
implying
that  
$\sgn\{
\newpartialmetric(\y,\y_0) \!-\! \newpartialmetric(\yy,\y_0)
\}
=\sgn(
\y - \yy
)$. \
It will then follow that 
$\Etamod(\yy,\y_0)$
is given by the right-hand side of (\ref{eq:Exch5}) with
$|\y - \yy|$ replaced by
$|\newpartialmetric(\y,\y_0) \!-\! \newpartialmetric(\yy,\y_0)|$, \
proving not only  
that $\Etamod \geqslant 0$  
but also that
$\Delta\uoverline(\y_{0}) > 0$ when
${\partial \qi(\yy)}/{\partial \yy} > 0$,
in the same way as below (\ref{eq:Exch5}).

Because
the reasoning below (\ref{eq:Exch5})
involves a finite neighborhood 
in the $\yy\antisliver\y\Sliver$ plane, it is enough to prove
monotonicity in the
interior of each block, more specifically that
$\partial\newpartialmetric(\y,\y_0)/\partial\y > 0$,
equivalently
${\partial^{2} \G(\y,\sliver\y_0)}/{\partial \y^{2}} > 0$, \
for $\y \neq \y_0$ \ and
$\y \neq \pm L$. \
It is here that we need the finiteness of $\LD$. \

Consider the graph of $\G(\y,\y_0)$
as a function of $\y$,
in each  
block  
$\y<\y_0$ and
$\y_0<\y$ separately.
From (\ref{eq:greens})
the second derivative satisfies
\begin{align}
\label{eq:curvature1}
\partial^2 \G(\y,\y_0) /\partial \y^2 &= \LD^{-2}\G(\y,\y_0)
&\mbox{for $\y \neq \y_0$}
\;. 
\end{align}
For finite $\LD$ 
the graph is therefore convex toward
the $\y$ axis
everywhere except at $\y = \y_0$
and $\y = \pm L$.  
Because the graph goes to zero
at both boundaries $\y = \pm L$,  
it can have 
only the one
extremum
at $\y=\y_0$.
The jump condition from (\ref{eq:greens}), 
\begin{align}
\left.\frac{\partial \G(\y,\y_0)}{\partial \y}\right|
^{\y=\y_{0+}}
_{\y=\y_{0-}}
= -1
\;,
\end{align}
ensures that the extremum is a maximum.
Therefore $\G(\y,\y_0)$ must be positive everywhere
apart from the boundaries $\y = \pm L$  
and therefore,
from
(\ref{eq:curvature1}),
\begin{align}
\label{eq:metricvalid}
\partial^{2} \G(\y,\sliver\y_0)/\partial \y^{2} > 0
\quad
\mbox{for all
$\,\y \;\neq\; \y_{0}$,
$\pm L$}
~.
\end{align}
This completes the proof.  
We have established that, for both
finite and infinite $L$,
the velocity change
$\Delta\uoverline(\y_{0})$
in the straightened-out  
jet core 
satisfies
\begin{empheq}[box=\framebox]{align} 
\label{eq:genjetresult}
\phantom{\Big|}  
\Delta\uoverline(\y_{0})>0
\;~\mbox{ if }~\:  
{\partial \qi(\yy)}/{\partial \yy}>0
~\mbox{ for all $\yy$\,,}  
\phantom{\Big|}  
\end{empheq} 
and vice versa,
for any nontrivial $\R$
that preserves
the eddy-transport barrier
at the jet core.

It is not clear
whether 
there is an alternative
proof analogous 
to
that of
Appendix~A. 
The counterpart
of the last term of
(\ref{eq:Killworth2})
no longer 
makes a vanishing contribution to
the counterpart of
(\ref{eq:DMhat3-casimir}).

\section{Beyond 
  the present models?}
\label{sec:beyond}

It might be thought that the beta-channel results should extend
to the full sphere
for finite as well as
for 
infinite $\LD$.  
However, such an extension would be far from
straightforward, if only because the standard \qg\ theory  
relies on $\LD$ being constant. 
Hence for finite $\LD$ the results are
   valid only 
to the extent that the beta channel is valid,  
namely, in a zone
   that is
narrow relative to the
planetary radius $\aaa$ and   
sufficiently far from the equator.
A remaining challenge, therefore,
is to make progress beyond the restrictions of
\qg\ theory and nondivergent barotropic theory,
$\LD=\infty$. \

Could there be an exact counterpart
to the basic theorem
(\ref{eq:basictheorem})?
The question makes sense 
at least for
thought-experiments
having a zonally symmetric final  
as well as initial state, 
with both states
in exact cyclostrophic balance.  
Then
PV invertibility tells us that there is an exact
counterpart to the question ``what is the sign of the absolute
\ahm\ change that results from generalized PV  
mixing?''
Here ``exact''  
indicates       
not only exact cyclostrophic balance but
also use of the exact (Rossby--Ertel) PV.

The conservation and
impermeability theorems satisfied by
the exact PV
\citep*{Haynes:1990b}
guarantee that the distinction between
generalized mixing and unmixing is still clear.
``Particles'' of
PV-substance or PV-charge (of either sign) can be thought of as
being
transported along isentropic surfaces, but never across them,
even when  
diabatic heating is significant.
Hence  
the upgradient transport involved in unmixing
means that PV-substance is transported against its
isentropic  
gradient.
Furthermore,
even though
the first-moment formula  
(\ref{eq:kelvin})
fails,
the total \aam\
is still well defined, and exactly defined.

An exact counterpart
to the basic theorem
(\ref{eq:basictheorem})
would therefore make sense  
as a conjecture.  
However, we have so far failed to prove any such exact theorem.
So the question remains open for now.  
The main technical obstacle appears to be
the nonlinearity
of the  
exact cyclostrophic 
PV  
inversion operator.

\section{Concluding remarks}   
\label{sec:conclu}

The basic theorem 
(\ref{eq:basictheorem})
proved here underlines the point
that, especially in problems of
jet formation and maintenance, as well as in ``beta-turbulence''
problems in general, it is advisable to consider
the
angular-momentum budget as well as the enstrophy and energy budgets.
The theorem
underlines
another fundamental point as well, namely
that
thought-experiments in which one imagines ``stirring'' the fluid
to mix the PV are not well defined until one specifies what is
doing the stirring.  
Artificial body  
forces
will in general cause some \emph{unmixing} of PV.
So too will immersed bodies such as
Welander's massless goldfish (P. B. Rhines 1971, personal
communication), which produce vortex quadrupoles
and are therefore capable of extending the range of PV values.
Indeed
massless goldfish, 
by definition,
cannot change
the 
\aam .
The goldfish
might
therefore produce profiles like those studied in DS08 and
illustrated in \fig \ref{fig:mix-examples}c above. 

Another
motivation for this work was to advance our understanding
of Jupiter's
weather layer.
An adequate representation of
what we observe on the real planet
will undoubtedly require a coupled model
of the weather layer and the underlying convection zone.
The convection zone is,
in turn,
bounded below by a strongly stratified transition
to metallic hydrogen,
as the pressure increases and
the proportion of ionized hydrogen atoms to neutral atoms and
H$_2$ molecules builds up with temperature.
It is likely that Richardson numbers
in the transition zone are enormous.
So it may well be that one  
can treat the transition zone as
a rigid but perfectly slippery boundary, whose only function
is to supply heat from below.

Our current aim is less ambitious, namely to isolate one aspect
of the coupling between the top of the convection zone and the
overlying
weather layer,
by making the simplifying assumption that the main effect of the
convection zone is to exert the fluctuating
form stress
required to catalyze PV mixing and jet formation.
For instance the form stress can be exerted via an artificial
``heaving topography'' $\topogfluct(\x,\y,t)$ 
acting as the forcing function on a \shw\ layer,
in place of the usual artificial body forces.  
Arguably, the addition of such quasi-topographic forcing  
might 
improve the realism
of simulations like that 
of \citet{Showman:2007}.
\citeauthor{Showman:2007}
also avoids using 
artificial body forces,
but assumes that
the sole effect 
of the 
convection zone 
is to 
produce
small-scale mass injections into the 
weather layer,
like thunderstorm anvils.

A further question is whether, with a
more natural and realistic forcing,
we can reach a statistically steady state
without having to invoke large-scale Rayleigh friction or
hypodiffusion, both of which are hardly natural assumptions
for a planet with no nearby solid surface.

These questions are as yet unanswered 
but we hope to make progress on them soon, 
through numerical experiments based on sophisticated
numerical codes that as far as possible respect the angular
momentum principle.

\begin{acknowledgment} 
We thank
Gavin Esler,
Kalvis Jansons,
Peter Rhines,
John Rogers,
Richard Scott,
Andrew Thompson,
Kraig Winters,
Wil\-liam Young,
and two reviewers
   for useful comments and discussion.
   RW's work is supported by a
   UK Science and Technology Facilities Council
   Research Studentship.
\end{acknowledgment}

\appendix[A]
\vspace{-1.5em}
        \section*{\begin{center}An alternative proof\end{center}}

\newcommand{\m}{m}
\newcommand{\foo}{A}
\newcommand{\zbar}{\breve{\eta}}  
\newcommand{\fbar}{\zbar^\dag}  
\newcommand{\sbar}{y^\dag}  

The connection to  
potential energy noted in
section~\ref{sec:ape}
suggests an alternative proof of the basic theorem,
via a mathematical route quite different
from that of section~\ref{sec:basictheorem}.
It
is motivated by  
positive-definite exact formulae for
po\-ten\-tial-energy changes  
that are already known
for purely advective rearrangements,  
of buoyancy  
(e.g.,
\break
\citeauthor{Holliday:1981}
\citeyear{Holliday:1981};
\citeauthor{Andrews:1981}
\citeyear{Andrews:1981};
\break
\citeauthor{Molemaker:2010}
\citeyear{Molemaker:2010};
\citeauthor{Roullet:2009}  
\citeyear{Roullet:2009}).  
These exact formulae are
now recognized as cases of the
energy--Casimir
and 
mo\-men\-tum--Casimir   
formulae arising in Hamiltonian models of disturbances to
nontrivial initial or background states
\citep[e.g.,][]{Arnold:1965,Shepherd:1993}.
The resulting proof of
(\ref{eq:basictheorem})
can be seen as  
a nontrivial 
generalization of the Hamiltonian theory, made possible by
the \mbox{$\R$-function} formalism.

For a purely advective rearrangement,  
the Hamiltonian  
formulae apply.   
In the \shw\ case, for instance,  
we have
\begin{equation}
\label{eq:Holliday}
\delM = - 
\densintd \dy \, 
 \foo(\y,\,\zbar) 
\end{equation}
where
$\zbar$  
is the latitudinal displacement of a fluid element
expressed as a function of its final latitude $\y$
rather than its initial latitude $\yy$, so that
$\zbar(\y)=\eta(\yy)=\y-\yy$,
and where
the function $\foo$ is defined by
\begin{equation}
\label{eq:Killworth}
\foo(\y,\,\zbar)
\,:=   
\int_0^{\zbar} d \fbar\, 
\frac{\partial \qi(\y-\fbar)}{~~\partial \y\quad}\; 
\fbar
~.
\end{equation}
It is only because of the
invertible mapping  
between the initial latitude
$\yy$ and final latitude $\y$ of a given fluid element,
in the purely advective case,  
that we can write the displacement
of that element  
as a function either
of~$\yy$ or of~$\y$.

We now show  
that
the $\R$-function formalism 
allows us to rederive
(\ref{eq:Holliday}) 
together with
its generalization beyond the Hamiltonian 
framework, as a single expression
\begin{equation}
\label{eq:DMhat3-casimir}
\framebox{$
\!\!
\phantom{\Bigg]^I}  
\delM
=
-\densintd\!\! \dy \intd\!\! \dyy \R(\yy,\y) \foo(\y,\,\y-\yy) 
~.
  ~~
$}
\end{equation}
First,  
we see by inspection 
that in the purely advective case,  
for which
$\R(\yy,\y) = \delta\left(\y - \yy - \zbar(\y) \right)$,
the expression
(\ref{eq:DMhat3-casimir})
does reproduce
(\ref{eq:Holliday}).
Second,
to see that (\ref{eq:DMhat3-casimir})  
is correct for a general \mbox{$\R$ function},
we rewrite
(\ref{eq:Killworth})
by substituting
$\sbar := y - \fbar$
and integrating by parts to obtain
\begin{align}
\label{eq:Killworth2}
\foo(\y,\,\y-\yy)
&\;=\;
- \qi(\yy)(\y-\yy)
- \int^{\yy}_{\y}\!\!\! d \sbar\, \qi(\sbar)  
~.
\end{align}
Now  
the 
last term of 
(\ref{eq:Killworth2})  
contributes nothing to 
(\ref{eq:DMhat3-casimir}).    
This is because it has the functional form  
$a(\y)-a(\yy)$.  
In virtue of the integral constraints
(\ref{eq:Rintconstraint-pvsubst})
and
(\ref{eq:Rintconstraint-wt-avg}),
the contribution to
(\ref{eq:DMhat3-casimir})
is  
$
\textstyle\intd\antisliver   
\textstyle\intd\!            
\sliver
\dy\dyy\,           
\R(\yy,\y)          
\left[             
   a(\y)-a(\yy)
\right]
=\sliver
$
$
\textstyle\intd\,
a(\y)
\dy
$
$
-
\textstyle\intd\,
a(\yy)
\dyy
=0
$,  
for any function $a(\cdot)$\,.

The definition
(\ref{eq:eta})
of the
average   
displacement
$\eta(\yy)$
can be rewritten using
(\ref{eq:Rintconstraint-pvsubst})
and
(\ref{eq:Rtilde})
as
\begin{align}
\nonumber
\eta(\yy)
&\,=
\intd \dy\, \Rd(\yy,\y) (\y-\yy)   
\\
\label{eq:eta2}
&\,=
\intd \dy\, \R(\yy,\y) (\y-\yy)
~.
\end{align}
Hence 
by substituting the first term of
(\ref{eq:Killworth2})
into
(\ref{eq:DMhat3-casimir}),
then using (\ref{eq:eta2}) to rewrite the result  
in terms of $\eta(\yy)$,
we see that 
(\ref{eq:DMhat3-casimir})
is equivalent to 
the original expression
(\ref{eq:DMhat2})
for $\delM$.
The basic theorem
(\ref{eq:basictheorem}) now
follows, because
(\ref{eq:Killworth}) shows that  
the function $\foo$
is positive definite whenever
$\partial\qi/\partial\y$
is positive, and  
negative definite whenever
$\partial\qi/\partial\y$
is negative.

\appendix[B]
\vspace{-1.5em}

\section*{\begin{center}Monotonizing $Q$\end{center}}

To see how to  obtain
monotonically increasing 
$\qm \Antisliver\left[ \yy;\qf\functional \right]$ 
from
the general 
$\qf(\xx,\yy)$
by advective rearrangement,  
one may
proceed as follows.

The function describing the
monotonic PV distribution $\q = \qm\Antisliver\left[ \y;\qf\functional \right]$ will have an 
inverse function 
$\y = \ym\Antisliver\left[ \q;\qf\functional \right]$.  
For a given $\q$ value,
all the fluid with
$\qf > \q$ will,
after rearrangement,
lie between
$\y = \ym\Antisliver\left[ \q;\qf\functional \right]$ and the northern
boundary $\y = L$.
Hence 
we may define
\begin{equation}
 \ym \Antisliver\left[\q ;  \qf \functional \right] 
  := 
  L - 
  \intd \!\! \dy \, 
  \overline{ {\mathcal{H}}\Antisliver\left\{\qf(\x,\y) - \q \right\}}
~,
\end{equation}
where 
$\mathcal{H}$ 
is the Heaviside step function
and
the
overbar again denotes averaging in $\x$.
The \rf\
representing the
advective rearrangement 
from 
$\qf(\xx,\yy)$
to
$\qm \Antisliver\left[ \yy;\qf\functional \right]$
is
\begin{equation}
\label{eq:rfwdappendix}
 \rfwd \Antisliver\left[\xx,\yy,\x,\y ; \qf \functional \right]
= 
\frac{
1
}{\textstyle \int\!\! \dxx}
\sliver
\delta  \big\{
          \y - \ym\Antisliver\left[ \qf(\xx,\yy) ; \qf\functional \right]
        \big\}
,
\end{equation}
where $\delta$ is the Dirac delta function.
\Eq(\ref{eq:rfwdappendix})
can be verified by
by substituting this $\rfwd$ into
(\ref{eq:rfwd}).

%%%%%%%%%%%%%%%%%%%%%%%%%%%%%%%%%%%%%%%%%%%%%%%%%%%%%%%%%%%%%%%%%%%%%
% REFERENCES
%%%%%%%%%%%%%%%%%%%%%%%%%%%%%%%%%%%%%%%%%%%%%%%%%%%%%%%%%%%%%%%%%%%%%

% Create a bibliography directory and place your .bib file there.

\bibliographystyle{ametsoc}
\bibliography{./bibliography/references}

\end{document}